# Efficient Attack Detection in IoT devices using Feature engineering-less machine learning

Arshiya Khan[1] and Chase Cotton[2]

[1]University of Delaware, Newark, USA
<arshiyak@udel.edu>
[2]University of Delaware, Newark, USA
<ccotton@udel.edu>

## Abstract

*Through the generalization of deep learning, the research community has addressed critical challenges in the network security domain, like malware identification and anomaly detection. However, they have yet to discuss deploying them on Internet of Things (IoT) devices for day-to-day operations. IoT devices are often limited in memory and processing power, rendering the compute-intensive deep learning environment unusable. This research proposes a way to overcome this barrier by bypassing feature engineering in the deep learning pipeline and using raw packet data as input. We introduce a feature engineering-less machine learning (ML) process to perform malware detection on IoT devices. Our proposed model," Feature engineering-less ML (FEL-ML)," is a lighter-weight detection algorithm that expends no extra computations on "engineered" features. It effectively accelerates the low-powered IoT edge. It is trained on unprocessed byte-streams of packets. Aside from providing better results, it is quicker than traditional feature-based methods. FEL-ML facilitates resource-sensitive network traffic security with the added benefit of eliminating the significant investment by subject matter experts in feature engineering.*

## 1. Introduction

Cyber Security experts have found pivotal features in network traffic, including packet captures (pcap). Data scientists have used them to fashion impressive models capable of differentiating malicious traffic from benign [1]. However, most network traffic is emitted over encrypted channels in the current scheme. This security measure has limited experts' ability to contrive meaningful features for machine learning (ML), which can soon become obsolete. This challenge has given birth to analyzing raw bytes to detect malicious behavior in internet flows.

In the Internet of Things (IoT) domain, devices are sensors that interact with the environment. They conditionally react to changes in the environment and exchange information over the internet about these changes. A typical example of an IoT device is a doorbell camera. We can check any activity at our door using the camera from anywhere on earth. It can also alert us of any break-ins. For an IoT device to operate, it must continuously communicate with its users, other devices, and servers without fail. This communication is done over the internet, where information travels in the form of packets. This flow of packets is commonly known as network traffic. For an IoT grid to function, the network traffic must be secure from cyber criminals. Intrusion Prevention Systems (IPS) (like Cisco Firewall and McAfee), and Intrusion Detection Systems (IDS) (like SolarWinds and Snort), can scan the network traffic and determine if they are secure or insecure. Historically, packets are validated using a pre-defined rule book. However, since the introduction of ML, many attempts have been made to detect insecure packets using ML. Both detection systems display substantial hardware constraints like processing power and sizeable memory requirements to store moving traffic and identify anomalous packets. The evolving world of smart sensors and IoT devices demands precision and speed simultaneously.

Researchers have used machine learning to detect traffic anomalies in several studies [2,3]. They have followed the standard pipeline of typical ML modeling. By extracting features from raw data, they have trained their models. In the network traffic specialization, they have manipulated values in the packets to create features. This conventional approach is not only ineffective over encrypted channels but is also task-sensitive. Attack scenarios on network traffic have evolved rapidly. Now, attackers can hit multiple levels of network architecture [4], rendering the task-sensitive approach defenseless. In 2020, a group of researchers published a survey [5] on the use of ML to enable security in IoT devices. They divided the survey into five sections dedicated to understanding the enormity of security challenges. The survey discussed current ML solutions like anomaly detection, attack evasion, and mitigation. However, it is only possible to use these methods if we already know the type of attack being launched. The survey also discussed the complexity of IoT networks and multi-level attacks [6] suffered by these devices. They can cause the IoT infrastructure to fail anywhere from the application layer to the physical layer. Such a complicated structure makes it impossible to design a defense model that can guard against all possible attack scenarios.

Therefore, it is imperative to leverage the generalized scope of ML, which can learn even minor irregularities from the dataset. Traditional attack-specific feature-based detection models can crop or manipulate the dataset in an unhelpful way.

The remaining paper is organized as follows. Section 2 is dedicated to a detailed review of contemporary work. We have discussed the use of deep learning models for network traffic in section 2.1 and their feasibility on IoT devices in Section 2.2. Section 3 goes into detail about our proposed methodology. It is divided into three parts. 3.1 discusses our proposed approach in detail. 3.2 talks about the dataset and preparation of the experiment, while 3.3 talks about our deep learning architecture. Section 4 explains the experiment results and performance comparison. Section 5 covers concluding remarks and future work.

## 2. BACKGROUND AND RELATED WORK

### 2.1. The advent of 1D-CNN in Malware Detection

The last decade has seen exemplary use of machine learning to classify network traffic. Studies based on these features usually use three approaches: 1) Port-based, 2) Deep packet inspection (DPI)-based, and 3) Behavior-based [7].

A port-based classifier classifies reliable ports as benign and unreliable ports malicious. However, this classification technique has been rendered unusable by employing port hiding techniques such as port camouflaging. Port randomization is also used to hide ports.

The port-based approach uses only headers to classify traffic; however, the **DPI-based** approach examines both header and the payload. The header is inspected to collect information about the sender application, and the payload signature is examined to ensure they are not invalid or blacklisted. Due to high computational costs, DPI-based is an inefficient method. Lastly, in the behavior-based approach, flow or session trends are observed. Features to train an ML model are fashioned from the statistical inferences of these trends.

Velan et al. [7] published a comprehensive survey of 26 papers in 2015. It investigated studies focused on encrypted traffic classification published between the period 2005 to 2014. Flow features were used in 12 of the 26 papers, packet features in 5, a combination of packet and flow features in 7, and other features in two.

Wang et al. [8] published another survey in 2019, which addressed the rise in raw packet content to train deep learning models. As more parts of the packets were encrypted, extracting features became difficult. Both headers and payload were concealed, forcing researchers to use raw data in the training set. From 2015 to 2018, 8 out of 12 publications used raw data from the packet for encrypted traffic classification, and two used both derivative features and raw data. One of the remaining two papers solely used packet-based features, while the other used flow-based features.

Wei Wang et al. [9] used raw bytes collected in groups of flows and sessions. This study employed the Representation Learning technique and converted raw bytes into images to create the training dataset. It is called USTC-TFC2016 [9], a private dataset pulled from the Stratosphere IPS Project Malware Dataset [10]. It used a two-dimensional convolutional neural network (2D-CNN) to classify the images. This experiment saved a small amount of computational power as it did not require feature extraction. However, it spent a lot on byte-to-image conversion and image training. The binary classifier (malware or benign) was 100% accurate, the 10-label classifier was 99.23% accurate, and the 20-label classifier was 99.17% accurate. The classifiers achieved an average accuracy of 99.41%. For the 10-label classifier, the lowest and the highest precision score were 90.7% and 100%, respectively. Similarly, the highest and the lowest recall score were 91.1% and 100%, respectively, for the 10-label classifier.

Wang et al. [11] conducted a similar study in 2017, where they used the ISCX VPN-non-VPN [12] dataset to perform encrypted traffic classification. This dataset contains six traffic types: chats, emails, file transfer, peer-to-peer (P2P), streaming, and Voice over Internet Protocol (VoIP) captured over both Virtual Private Network (VPN) and non-VPN settings. This study performed binary classification on VPN and non-VPN traffic and multi-label classification on the six traffic types. It used raw bytes of flows and sessions of the ISCX dataset. To train on raw bytes, it used a one-dimensional convolutional neural network (1D-CNN). The 2-label (binary) classification achieved 100% precision, and the 6-label classification achieved 85.5% accuracy. Six traffic types were branched into 12 classes using VPN as a factor (VPN and non-VPN traffic). The 12-label classification achieved 85.8% accuracy. This study compared their results with [9], which used 2D-CNN. In a series of 4 experiments, 1D-CNN outperformed 2D-CNN by as much as 2.51%.

DeepPacket [13], introduced in 2017, is a deep learning framework for network traffic characterization and application identification. It employed two deep learning techniques: a) stacked autoencoder (SAE) neural network and b) 1D-CNN. Both were trained on ISCX VPN-non-VPN [12] dataset. The traffic characterization task classified VPN and non-VPN traffic. With SAE, the binary classifier's average precision and recall were 92%. However, with 1D-CNN, the same classification resulted in average precision of 94% and an average recall of 93%. The application identification task classified packets into these six applications: chats, emails, file transfer, peer-to-peer (P2P), streaming, and Voice over Internet Protocol (VoIP). With the SAE classifier, the average precision of the 6-label classification was 96%, and the average recall was 95%. With the 1D-CNN classifier, the average precision and recall were 98%.

In 2020, Rezaei and Liu [14] employed multi-task learning to classify network traffic. It divided the classification task into two tasks: a) bandwidth requirement and b) predicting the duration of traffic flow. It used two datasets for training: ISCX VPN-non-VPN and QUIC [15]. The multi-task learning model was trained on a 1D-CNN using these three time-series features: i) packet length, ii) inter-arrival time, and iii) direction of the traffic. On the ISCX dataset [12], their model classified traffic with 80.67% accuracy. The same experiment classified bandwidth requirement and prediction of flow duration with 88.67% and 90% accuracies, respectively. On the QUIC dataset [15], the model classified traffic with 94.67% accuracy. In this experiment, bandwidth requirement and flow duration prediction attained 90.67% and 91.33% accuracies, respectively.

Table 1. Related Works of Network Traffic ML models

| Work | DL Technique | Model Input | Dataset | ML Task | Accuracy | Year |
|------|--------------|-------------|---------|---------|----------|------|
| [9] | 2D-CNN | images | [9] | Malware detection | 99.23% | 2017 |
| [11] | 1D-CNN | bytes | [12] | Traffic characterization | 100% | 2017 |
| [13] | 1D-CNN + SAE | bytes | [12] | Traffic characterization | 98% | 2017 |
| [14] | 1D-CNN | time series | [12, 15] | Traffic characterization | 94.67% | 2020 |
| [17] | 1D-CNN + LSTM | raw bytes | [9] | Malware detection | 98.6% | 2020 |

Huang et al. [16] published a survey in 2019. It was based on deep learning use cases in the time-series domain. Cyber Security was one of the emergent real-world disciplines in the time-series domain, along with health, finance, and transportation. To achieve this conclusion, the paper analyzed topics such as traffic classification, anomaly detection, and malware identification.

Marin, Casas, and Capdehourat [17] published a study in 2020 aiming to remove engineered features, thus eliminating the need for domain experts. They reintroduced two types of malware detection approaches using raw packet content: raw packets and raw flows. They used raw byte-streams from pcaps to train their deep learning model to achieve this goal. Their ML model was a combination of a Long short-term memory (LSTM) network and a 1D-CNN. The raw byte-stream of flow performed (98.6% accuracy) better than the byte-stream of a packet (77.6% accuracy). A comparative experiment was performed between traditional feature-based and raw byte-based models. For the traditional model, they used a random forest (RF) and trained it on 200 in-flow features. Both raw byte-based models performed better than RF.

Discussion of the previous works in this section has revealed two significant points:

1. They indicate the advent of 1D-CNN in the network traffic characterization and malware detection domain. It is also evident in Table 1, which displays several publications of our study. The table is arranged chronologically and shows a clear trend of ML-based network traffic classifiers preferring 1D-CNN over other techniques. In addition to its superior performance, 1D-CNN has the advantage of preserving the time-series nature of network traffic. Inspired by its effectiveness, we have used it as the modeling baseline of our study.

2. Studies also suggest that in recent years, malware detection models have moved from engineered feature ML to non-engineered feature ML using raw data. However, these models may only be applicable to some network traffic use cases. It is practically impossible to employ complex models on memory-constrained devices like IoT. More extensive models like LSTMs and 2D-CNNs need several preprocessing steps to extract raw data and train a model on them. In the next section, we will discuss the use of ML practices in the IoT environment.

**2.2. ML Techniques for Malware Detection in IoT environment**

Traditional appliances, devices, and machines have moved to smart sensor technologies in the last few decades. In addition to routers and IPSs, these technologies are also a component of the IoT. Uninterrupted internet access makes them susceptible to malicious attacks, which may result in malfunction or failure. Several studies have tried to find security solutions for these IoT devices.

In 2019, Shouran, Ashari, and Priyambodo [18] introduced a straightforward way of detecting threats in IoT devices. It classified every device interaction with the internet into Low, Medium, and High impact. The classification criteria included compromise in Confidentiality, Integrity, and Authenticity (CIA).

Another research in 2019 [19] introduced an elaborate IDS for IoT devices. However, the IDS was feature-based with three layers of ML algorithms.

Later in 2020, a paper published by Vinayakumar et al. [20] developed a two-level deep learning model which discriminates malicious traffic from benign. The first level detected the most frequent DNS queries, and the second level used a domain generation algorithm (DGA) to detect illegal domains.

Sriram et al. [21] presented their deep learning ML system, which used network flows to find statistical features. Two datasets were used and compared over different ML modeling techniques like logistic regression, random forest, and LSTM. [17] also presented a 1D-CNN model for botnet detection on IoT devices. Details of their work are discussed in section 2.1.

Table 2. Related Work in IoT domain

| Work | IoT Malware detection Technique | Task | Year |
|---|---|---|---|
| [18] | Non-ML (Rule-based) | Malware detection | 2019 |
| [19] | ML-based IDS | Traffic characterization | 2019 |
| [20] | ML based DNS categorization | Malware detection | 2020 |
| [21] | Ensemble of ML models | Botnet detection | 2020 |
| [17] | 1D-CNN | Malware detection | 2020 |

Table 2 shows a task-based analysis of research works published on detecting malware on IoT devices. Most of the methods presented here require multiple steps to perform one task. This approach is not suitable for the IoT environment.

Our contribution is as follows:

1. Outright elimination of 'engineered features' that require additional computation. We introduce a network traffic classification system in favor of more lightweight features which come directly from the input data. As a result, it does not require domain-based expertise to perform feature engineering.

2. Increase the speed of classification by using 1D-CNN to train the deep learning model. The model will consume less memory during both the training and testing phases allowing it to be deployed on IoT devices.

## 3. METHODOLOGY

### 3.1. Feature engineering-less ML

IoT devices at the edge, like voice-based virtual assistants (e.g., Amazon Echo), smart appliances, and routers, must react to change at a very high speed. Consequently, they have to validate incoming traffic in real-time. They also have a limited battery life to support these unremitting transactions. In an internet-dependent environment like this, security from cyber attacks is non-negotiable; nevertheless, it can become an overhead. It can be in various forms, like malware recognition, anomaly detection, or behavior classification. Detecting cyber attacks using ML has shown promising results in the past [22, 23]. However, their deployment on IoT devices is unrealistic, as they involve computationally extensive feature engineering and require deep classifiers for precision. Feature engineering is a step-by-step process that incorporates: i) extracting desired elements from raw data, ii) cleaning and converting them into features, iii) standardizing features, and iv) aggregating them to be used in the classifier. For a time sensitive IoT environment, it is a complex and time-consuming operation. We cannot rely on traditional methods to make ML adequate on IoT devices. In this study, we propose skipping feature engineering and developing an IoT-friendly deep learning technique called Feature Engineering-less ML.

Feature engineering-less ML or FEL-ML is a product of the featureless modeling technique. Feature engineering-less modeling is machine learning without feature engineering. It is a lighter-weight detection algorithm where no extra computations are expended to compute 'engineered features,' resulting in an adequate acceleration to the low-powered IoT. We eliminate the feature extraction and processing step from the ML pipeline in FEL-ML. We store the streams of packets that arrive at an IoT device in their raw state. We create our training dataset by converting the raw streams to raw byte streams. The rest of the deep learning process remains the same. The advantages of FEL-ML are that it conserves the properties of a stream of bytes and saves time during the process. Omitting the feature extraction and generation step makes both model training and testing efficient.

Feature engineering-less modeling eliminates two significant IoT overheads: *computation cost* and *human cost*. Human cost involves using technical expertise to collect and clean traffic data. It also employs domain expertise to manipulate them into meaningful representatives of the dataset. Computation cost incorporates the computational overhead a device endures toward statistical operations at a device's central processing unit (CPU). Further down the ML pipeline, tensor-based computations like convolution, batch processing, and running multiple epochs contribute to computational overhead.

We have represented network traffic in three views: 1) Raw session, 2) Raw flow, and 3) Raw packet. Based on these three representations, we have developed three unique ML models. In the end, we will compare their performance to determine the most suitable model.

Raw session: In an IoT network, a session is a bi-directional stream of communication be- tween two devices. All packets in a session share these 5-tuple attributes: a) source IP address, b) source port, c) destination IP address, d) destination port, and e) transport protocol present in each packet header. In this representation, we split the pcap files into unique sessions based on the 5-tuple. We used MIT's *pcapsplitter* tool [24]. A pcap file records the live traffic stream on a device. Each individual session stream forms a unique component in the ground truth.

Raw flow: A flow is similar to a session, except that they are unidirectional. A flow is a batch of packets going from one device to another. We used SplitCap [25] to prepare our dataset in this representation. SplitCap splits pcap files into flows. A sample in the ground truth is a byte stream of a flow.

Raw packet: A pcap file consists of one or more packets. In this representation, we used the byte stream of each packet. A single packet in byte format is an individual ground truth component.

In a similar study [17], raw bytes were used to train the classifier. However, their experiment only included packets and sessions. Further, they have not discussed any memory or system metrics to demonstrate its efficiency after eliminating feature engineering. Their model constituted 1D-CNN and LSTM layers, i.e., a deep model. Deeper models are more complex and can result in overfitting. As a result, we have used a smaller but effective neural network to overcome this obstacle.

### 3.2. Experimental Design

We have used the Aposemat IoT-23 dataset [26] to perform experiments for this study. It is a labeled dataset captured from 2018 to 2019 on IoT devices and published in 2020 by the

Stratosphere Research Lab. Several studies have either discussed it in their work or used it in their model. Bobrovnikova et al. [27] used this dataset to perform botnet detection in 2020. They extracted features to curate numerical features for classification. They attained 98% accuracy with support vector machines (SVM) [28]. Blaise et al. [29] have mentioned that this dataset is similar to what they required but not pertinent to their experiments. This dataset also contained *conn.log.labeled* file generated by employing Zeek on this dataset. Stoian [30] used numerical and categorical features from this file. They trained a Random Forest classifier [31] with this dataset and attained 100% accuracy.

IoT-23 has 23 sets of scenarios, out of which three are benign and 20 are malicious. The benign dataset is collected from three IoT devices: i) Amazon Echo, ii) Somfy smart door lock, and iii) Philips hue LED lamp. These devices are then infected with an assortment of botnet attacks executed in a simulation environment forming the malicious dataset.

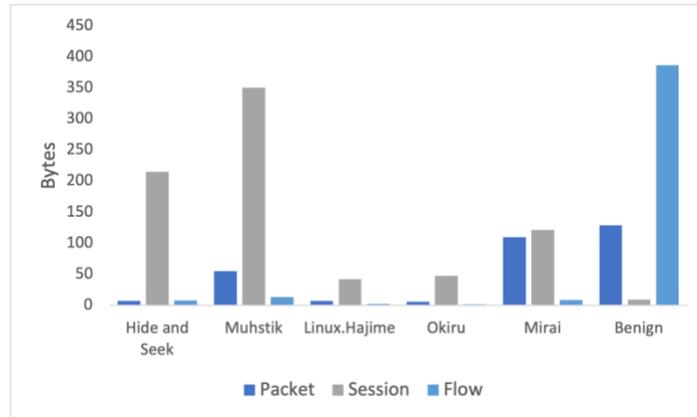

Figure 1. Byte Distribution

IoT-23 contains raw pcap files and Zeek files of each scenario. For our experiments, we used the pcap files as our training dataset. We integrated the three benign scenarios into one extensive dataset. The malicious scenarios are prepared in a simulation environment where the three devices are infected by 20 unique botnet attacks. Pcap files of several of these botnets range between hundreds of GBs. To demonstrate the fitness of our proposed model on limited-memory devices, we used attack scenarios with smaller sizes. We have selected five out of 20 attacks: i) Hide and Seek, ii) Muhstik, and iii) Linux.Hajime, iv) Okiru, and v) Mirai. Byte distribution of the five infected and one benign traffic data is shown in Figure 1. Using binary classification, we distinguished between benign and malicious traffic, while multi-label classification distinguished between botnet traffic.

As mentioned in section 3.1, we represented this dataset in three unique views: session, flow, and packet. In order to use raw bytes, we converted all traffic representations into a hexadecimal format using tshark [32]. As we take a deep dive into this experiment, we named these three experiments as follows:

*ExpS*: The experiment used the session representation of the dataset in the hexadecimal format. In the training data, each session is represented as one row of a byte stream.

*ExpF*: The experiment used the flow representation of the dataset in the hexadecimal format. In the training data, each flow is represented as a byte stream.

*ExpP*: The experiment used the packet representation of the dataset in the hexadecimal format. Each pcap instance is a sample in the labeled training set.

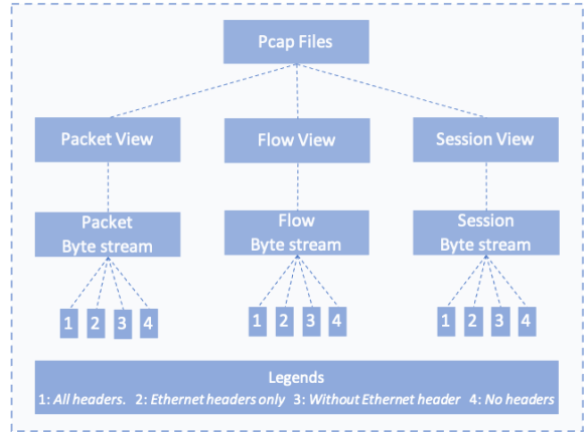

Figure 2. Training data generation

Studies [17] conducted in the past on raw byte streams have an additional step to remove ethernet and TCP/IP headers from their dataset. We experimented with the use of headers to achieve our goal of diminishing computational overhead further. We reformulated the dataset into four unique categories for each ExpS, ExpF, and ExpP. The first category included all headers in the packet along with the payload. We called this category: *All headers*. In the next category, we kept the ethernet headers and discarded the IP headers; hence we called this category *Ethernet headers only*. Intuitively, in the third category, we removed ethernet headers from the packet and called this category: *Without ethernet*. Ultimately, we dropped both ethernet and IPv4 headers from the packet. It was called the *No headers* category. As a result, all three representations of traffic (session, flow, and packet) were split into these four categories, and each category was then trained separately. As shown in Figure 2, there were 12 different experiments in our study. All categories in each representation had the same size.

### 3.3. DL Architecture

At a device, packets arrive as instances of data distributed over time which puts our dataset into the time-series domain. Until recent years 1D-CNNs have been primarily used in the Natural Language Processing (NLP) models [33]. However, they have also successfully classified time series data [34]. This study aims to develop a small neural network that can be installed on resource constraint devices and detect malicious streams of packets. Since tensor computations in 1D-CNN take less space than 2D-CNN and other ML methods, it has motivated us to use this lightweight neural network as our machine learning model. A smaller neural network size will increase its practicability on IoT devices.

As shown in Figure 3, our smaller neural network comprises two 1D-CNN layers, a Maxpooling layer, a Dropout layer, and a Dense layer. We started with the first convolution layer that performed convolution with a kernel size of 64 and a stride of 3 on the input vectorized byte stream. Small values of these parameters reduced the complexity of the model resulting in less overfitting. The maxpooling layer was sandwiched between the two convolution layers. It performed a 5-to-1 pooling operation to reduce the output tensor size from the upper layer without losing significant properties. The second convolution layer was placed after the maxpooling layer and performed the same job as the first convolution layer. It also had the same environmental controls. Next, we used the dropout layer with a drop rate of 0.5 to reduce validation loss. In the end, we added a fully connected Dense layer that helped the model learn any non-linear relationship between features.

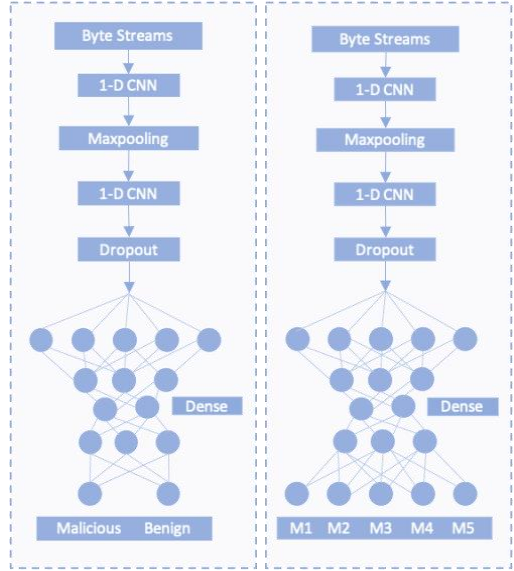

Figure 3. DL Architecture

The left section of Figure 3 depicts binary classification. It used the binary loss as the loss function and the softmax function for activation. On the other hand, multi-label classification, as depicted in the right section of the figure, used categorical cross-entropy as a loss function and a sigmoid function for activation. The binary classifier performed classification between benign and malicious traffic. We trained the multi-label model to classify the five botnet categories mentioned in section 3.2. The neural network trained over 60,000 hyperparameters in batches of 32 using Keras [35] for 50 epochs. For efficiency, we also used a checkpoint function to store the best model whenever encountered during training. It enables TensorFlow [36] to stop training when it achieves the best possible value of the evaluation metric, which is "Accuracy," in this case. When the hyperparameters are suboptimal, the resultant model becomes complex, leading to overfitting and high validation loss. This will result in more power usage and potentially incorrect classification.

We trained on Nvidia GeForce GTX 1060 GPU with a 12GB Ubuntu 16.04 server on an x86 architecture.

## 4. EVALUATION

### 4.1. Evaluation Metrics

Experiments were evaluated on two metrics: accuracy and f-1 score, as shown in Figure 4. Accuracy is the percentage of correct results from the total results produced by the model. It is calculated on both training and validation data. The f-1 score is the weighted average of precision and recall values on the validation data. f-1 score is also suitable for datasets with uneven distribution which makes it suitable for our experiments.

$$Accuracy = \frac{Correct\ predictions}{Total\ predictions}$$

$$f1 - score = \frac{2}{\frac{1}{precision} + \frac{1}{recall}}$$

$$precision = \frac{True\ positive}{True\ positive + False\ positive} \quad recall = \frac{True\ positive}{True\ positive + False\ negative}$$

Figure 4. Evaluation Metrics

## 4.2. Experimental evaluations

Our experiments indicate that binary classification achieved better accuracy the multi-label. First, we will discuss binary classification outcomes between malicious and benign traffic. Tables 3 and 4 display evaluation metrics in each ExpP, ExpS, and ExpF.

Table 3. Binary Performance on IoT-23 dataset

| Representation | Header | Accuracy | f1-score |
|---|---|---|---|
| ExpS | All headers | 1.00 | 0.97 |
| | Only Eth | 1.00 | 0.96 |
| | Without Eth | 1.00 | 0.96 |
| | No headers | 1.00 | 0.94 |
| ExpF | All headers | 1.00 | 0.97 |
| | Only Eth | 1.00 | 0.93 |
| | Without Eth | 0.97 | 0.96 |
| | No headers | 0.99 | 1.00 |
| ExpP | All headers | 1.00 | 0.96 |
| | Only Eth | 0.98 | 0.96 |
| | Without Eth | 0.98 | 0.97 |
| | No headers | 0.99 | 0.95 |

Table 4. Multi-label Performance on IoT-23 dataset

| Representation | Header | Accuracy | f1-score |
|---|---|---|---|
| ExpS | All headers | 0.99 | 0.96 |
| | Only Eth | 0.94 | 0.93 |
| | Without Eth | 0.84 | 0.92 |
| | No headers | 0.96 | 0.92 |
| ExpF | All headers | 0.93 | 0.92 |
| | Only Eth | 0.72 | 0.85 |
| | Without Eth | 0.79 | 0.91 |
| | No headers | 0.91 | 0.90 |
| ExpP | All headers | 0.97 | 0.93 |
| | Only Eth | 0.98 | 0.93 |
| | Without Eth | 0.74 | 0.80 |
| | No headers | 0.98 | 0.93 |

In pcap experiment ExpP, the *"no-header"* category achieved the maximum accuracy of 99%. However, in the *"all-headers"* category, ExpP achieved 97% accuracy, only 0.02% below the highest but gave the highest 99% f-1 score.

In ExpS, binary classification achieved 100% accuracy in every header category. It also achieved the highest f-1 score of 97% in the *"all-headers"* category.

In ExpF, the *"all-headers"* category again achieved 100% accuracy alongside the "only-ethernet" category. In contrast, ExpF achieved a 100% f-1 score "no-header" category, while the *"all-headers"* category achieved only a 97% f-1 score.

We now switch our attention to multi-label classification between five botnet attack scenarios. Table 4 shows the performance of the evaluation metrics in all categories: ExpP, ExpS, and ExpF. Overall, the 5-label classification achieved a maximum accuracy of 99% and an f-1 score of 96% in the ExpS session scenario.

In ExpP, accuracy was 98%, and the f-1 score was 93% f-1 score in the "no header" category. However, the *"all-headers"* category was only 0.01% behind with a 97% accuracy and the same 93% f-1 score.

In ExpS, the *"all-headers"* category achieved the highest accuracy of 99% with a 96% f-1 score.

In ExpF, the *"all-headers"* category again achieved the highest accuracy of 93%, along with the highest f-1 score of 92%.

Table 5. Binary comparison between IoT-23 and ETF-IoT Performance

| Representation | Header | IoT-23 f-1 | ETF-IoT f-1 |
| --- | --- | --- | --- |
| ExpS | All headers | 0.97 | 0.89 |
| | Only Eth | 0.96 | 0.87 |
| | Without Eth | 0.96 | 0.87 |
| | No headers | 0.94 | 0.90 |
| ExpF | All headers | 0.97 | 0.87 |
| | Only Eth | 0.93 | 0.86 |
| | Without Eth | 0.96 | 0.87 |
| | No headers | 1.00 | 0.98 |
| ExpP | All headers | 0.96 | 0.88 |
| | Only Eth | 0.96 | 0.89 |
| | Without Eth | 0.97 | 0.89 |
| | No headers | 0.95 | 0.89 |

Evidently, session representation or ExpS performed better in binary and multi-label classifications. Overall, accuracy was highest when the dataset included all headers. Accuracy monotonically decreased when either header was removed. However, it recovered when there were no headers. The f-1 score was always the highest when all headers were included, with one exception in multi-label ExpF.

This study shows that headers significantly influence the precision of anomaly detection models. Unlike the custom of cropping them out of the training set [17] we achieve better results by incorporating them into the training set.

## 4.3. Comparison with another Dataset

In this section, we compare the results of our model trained on another dataset named ETF IoT Botnet [37]. ETF is the newest publicly available botnet dataset. It has 42 malicious botnet attack scenarios collected on RaspberryPi devices and two benign scenarios.

Tables 5 and 6 show lateral comparisons between both datasets. We have shown f-1 score comparisons. The results differ case by case depending on the placement of the header.

Table 6. Multi-label comparison between IoT-23 and ETF-IoT Performance

| Representation | Header | IoT-23 f-1 | ETF-IoT f-1 |
|---|---|---|---|
| ExpS | All headers | 0.96 | 0.95 |
| | Only Eth | 0.93 | 0.93 |
| | Without Eth | 0.92 | 0.94 |
| | No headers | 0.92 | 0.93 |
| ExpF | All headers | 0.93 | 0.96 |
| | Only Eth | 0.85 | 0.95 |
| | Without Eth | 0.91 | 0.96 |
| | No headers | 0.90 | 0.94 |
| ExpP | All headers | 0.97 | 0.96 |
| | Only Eth | 0.93 | 0.95 |
| | Without Eth | 0.80 | 0.96 |
| | No headers | 0.93 | 0.93 |

Figures 5 and 6 show a comparative analysis of f-1 scores between the two datasets. Both datasets show similar results in all categories with a few exceptions. Noticeably, the *"all headers"* category did not show any change in trend. f-1 scores of this model are consistent with IoT-23, which strengthens the claim of FEL-ML's usability for malware detection. As the obvious next step, we have tested our model's feasibility to be deployed on IoT devices.

We selected five botnet classes using the same technique we used in IoT-23. They are: 1) 666, 2) SNOOPY, 3) arm7.idopoc2, 4) z3hir arm7, and 5) arm7l 1. We also used the same tools and scripts to extract session, flow, and packet representation from pcap files. We trained on the same GPU setting. All parameters of the 1D-CNN were also the same.

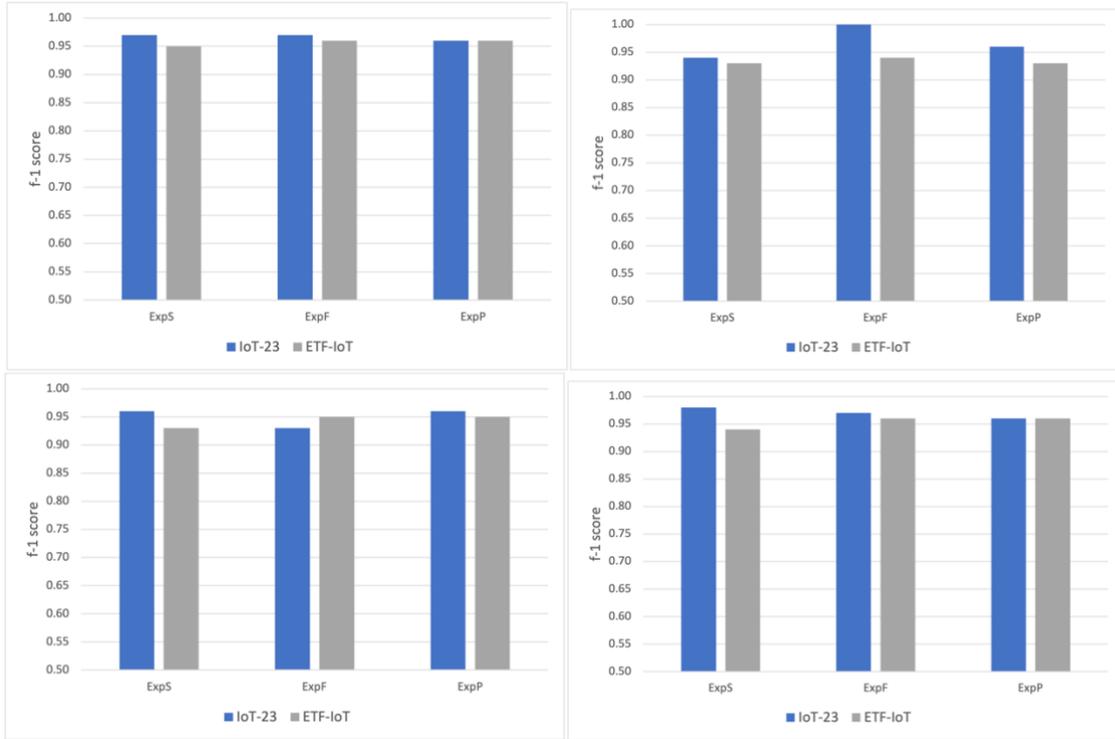

Figure 5 f-1 score comparison of Binary Classification (a) All headers (b) No headers (c) With Ethernet headers (d) Without Ethernet headers

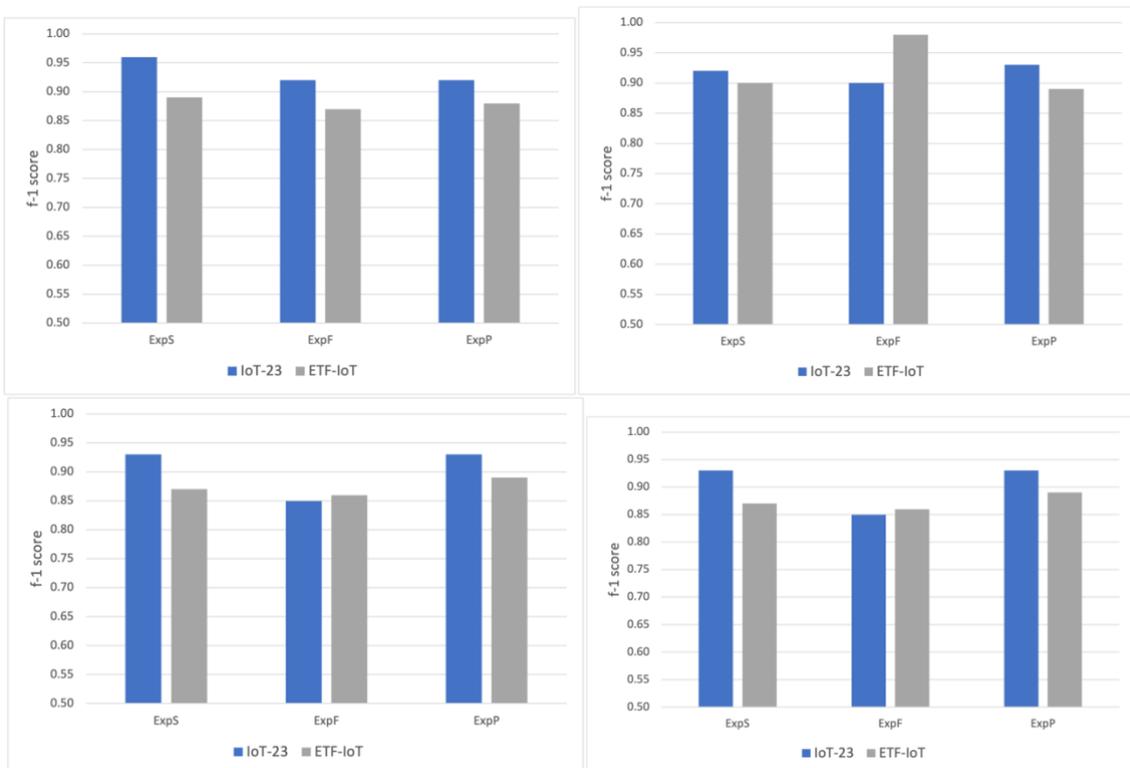

Figure 6 f-1 score comparison of Multi Classification (a) All headers (b) No headers (c) With Ethernet headers (d) Without Ethernet headers

## 4.4. Applicability in IoT scenario

These devices are cost-sensitive, resulting in lower-performing CPUs and less memory, demanding much lower-cost detection schemes. We couldn't find a study that addresses this issue. It prompted us to develop a faster ML model than the existing standards. In this section, we compare our FEL-ML's performance with an existing feature-based ML model. Both models were trained and tested on the same GPU-enabled architecture.

Since our focus is on botnet attack detection, for this experiment, we selected an ML model called n-BaIoT [2], which detects botnet attacks on IoT devices. We used the [38] GitHub repository to reproduce their model and train on the dataset used in the original study. We extracted 115 numerical features from the dataset, as mentioned in the paper. We performed binary classification between benign and malicious traffic to compare model performances. We used three performance metrics in this experiment:

1. Time consumed on testing the dataset, measured in seconds,

2. System time,

3. CPU utilization.

We measured 1 using the *"time"* function in python. We recorded 2 using the Linux time function. We measured 3 using the *"perf"* tool on Ubuntu [39]. perf is a profiling tool that provides kernel-level information about a program when it executes.

Table 6. Performance applicability for IoT devices

| Model | Accuracy | Time elapsed (sec) | System time (sec) | CPU utilization (max:2) |
|---|---|---|---|---|
| Binary ExpS | 1.00 | 2.813 | 0.71 | 1.171 |
| Binary ExpF | 1.00 | 7.269 | 0.82 | 0.513 |
| Binary ExpP | 1.00 | 29.626 | 2.42 | 1.350 |
| Binary n-BaIoT | 0.99 | 22.877 | 1.30 | 1.238 |

All the binary classifications of our model used the *"all-headers"* category of the dataset. As displayed in Table 7, the n-BaIoT model trained on 115 features took 22.877 seconds to perform testing. However, its accuracy remained at 99.96%. Our featureless model performed better in session and flow (ExpS and ExpF) representations, where it used less Testing Time compared to the feature-based model. Testing time of only ExpP pcap representation took 6.746 seconds more than the n-BaIoT model. Similar trends were seen in system time. Similarly, session and flow utilize less CPU compared to feature engineered models.

## 5. CONCLUSION

Contrary to traditional ML methodologies, FEL-ML does not require the complex processing power deemed necessary. With the ease of implementation, more industrial domains can now include it in their day-to-day operations. Security of IoT devices is one such domain. With IoT devices running on battery power, more accurate results can be achieved with less computational overhead by training raw bytes on 1D-CNN. As well as identifying anomalies in traffic with 100% accuracy, this methodology is able to identify their types with 99% accuracy. Previous works on this topic have scraped headers from their training set. However, our experiment compares models trained with and without headers. This extensive experiment reinforces our argument that feature engineering and removing headers from packets is a step in the traffic classification process that is unnecessary.

As evident from Table 7, one challenge faced by our algorithm is that pcap representation could perform better than flow and session. Flow and session require additional overhead in consolidation before feature engineering. The next step of this research will attempt to discover simpler ML systems that are efficient for direct packet captures. Speed of detection and accuracy are of utmost importance to performing more granular detection of malware on IoT devices.

**Authors**

Arshiya Khan is currently pursuing her Ph.D. in Electrical and Computer Engineering (Cybersecurity) at the University of Delaware, Newark, DE, USA. Her areas of interest include network security, artificial general intelligence, and fair machine learning. She wrote her M.S. thesis on feature taxonomy of network traffic for machine learning algorithms.

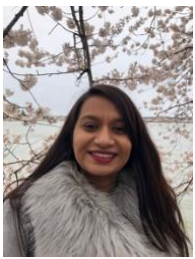

Over the past 35 years, Chase Cotton (Ph.D. EE, UD, 1984; BS ME, UT Austin, 1975, CISSP) has held a variety of research, development, and engineering roles, mostly in telecommunications. In both the corporate and academic worlds, he has been involved in computer, communications, and security research in roles including communication carrier executive, product manager, consultant, and educator for the technologies used in Internet and data services.

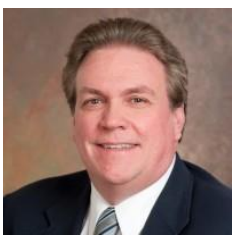

Beginning in the mid-1980 Dr. Cotton's communications research in Bellcore's Applied Research Area involved creating new algorithms and methods in bridging, multicast, many forms of packet-based applications including voice & video, traffic monitoring, transport protocols, custom VLSI for communications (protocol engines and Content Addressable Memories), and Gigabit networking.

In the mid-1990s, as the commercial Internet began to blossom, he transitioned to assist carriers worldwide as they started their Internet businesses, including Internet Service Providers (ISPs), hosting and web services, and the first large scale commercial deployment of Digital Subscriber Line (DSL) for consumer broadband services. In 2000, Dr. Cotton assumed research, planning, and engineering for Sprint's global Tier 1 Internet provider, SprintLink, expanding and evolving the network significantly during his 8-year tenure. At Sprint, his activities include leading a team that enabled infrastructure for the first large-scale collection and analysis of Tier 1 backbone traffic and twice set the Internet 2 Land Speed World Record on a commercial production network.

Since 2008, Dr. Cotton has been at the University of Delaware in the Department of Electrical and Computer Engineering, initially as a visiting scholar, and later as a Senior Scientist, Professor of Practice, and Director of Delaware's Center for Intelligent CyberSecurity (CICS). His research interests include cybersecurity and high-availability software systems with funding drawn from the NSF, ARL, U.S. Army C5ISR, JPMorgan Chase, and other industrial sponsors. As Director, Cybersecurity Minor & MS Programs, he currently is involved in the ongoing development of a multi-faceted educational initiative at UD, where he is developing new security courses and degree programs, including a minor, campus and online graduate Master's degrees, and Graduate Certificates in Cybersecurity.

Dr. Cotton currently consults on communications and Internet architectures, software, and security issues for many carriers and equipment vendors worldwide.